\documentstyle[12pt,fleqn]{article}
\begin{document}
\baselineskip= 22 truept
\def\be{\begin{equation}}
\def\ee{\end{equation}}
\def\bea{\begin{eqnarray}}
\def\eea{\end{eqnarray}}
\def\pa{\partial}
\begin{titlepage}
\begin{flushright}
IP/BBSR/96-13\\
hep-th/9602144 \\
\end{flushright}
\vspace{1cm}\begin{center} {\large \bf M-Theory
on Orientifolds of $K_3\times S^1$}\\ \vspace{1cm} {\bf 
Alok Kumar and Koushik Ray
 }\\ Institute of Physics, \\Bhubaneswar 751 005,
INDIA \\ email: kumar, koushik@iopb.ernet.in \\ 
\today
\end{center} 
\thispagestyle{empty}
\vskip 4cm
\begin{abstract}
We present several Orientifolds of M-Theory on $K_3\times S^1$ 
by additional projections with respect to 
the finite abelian automorphism groups of $K_3$. 
The resulting models correspond to 
anomaly free theories in six dimensions. We construct
explicit examples which can be interpreted as models 
with eight, four, two and one vector
multiplets and $N=1$ supersymmetry in six dimensions. 
\end{abstract}
\vfil
\end{titlepage}
\eject

$M$-Theory\cite{mth,mth2,maha,horav,mukhi,wit2,mtdual,davis,sen}, 
believed to be a candidate for the unification of
all string theories, is at present a focus of attention. At low
energies, this theory is represented by the eleven-dimensional
supergravity. It was already shown earlier, that the
eleven-dimensional supergravity is the strong coupling limit of
the ten dimensional type IIA string theory\cite{mth}. 
Recent interest
in the subject was generated by the fact that the
compactification of these theories to ten dimensions on an
orientifold\cite{sagnotti,leigh,polch}
of $S^1$ gave rise to the $E_8\times E_8$ heterotic
string theory in ten dimensions\cite{horav}. 
In proving this equivalence, the
anomaly cancellations for the ten dimensional $N=1$
supersymmetric theories plays an important role\cite{alvarez}. 
This is due to
the fact that, in the absence of a complete knowledge of this theory,
an explicit construction of the orientifolds is not possible. 
More recently, compactifications of $M$-Theory to six
dimensions have also been studied. Its 
$T^5/Z_2$ orientifold gave
rise to an anomaly free $N=2$ supersymmetric theory in six dimensions
with 21 tensor multiplets\cite{mukhi,wit2}. 
Once again the requirements of the
anomaly cancellation\cite{anomaly} 
played a major role in determining the complete 
spectrum of this theory. It was shown that the fixed
points of the torus degenerate into pairs. As a result, the
twisted sectors contribute only sixteen extra tensor multiplets,
instead of thirty two. 

The orientifolds of type II string theories have also been examined
\cite{vawit,kumar,dabol}. In \cite{dabol}
construction of a new chiral string theory in six
dimesnions, through an orientifold compactification of the type
IIB strings on $K_3$, has been presented. 
It was shown that such a compactification
gives rise to an $N=1$ supersymmetric
theory in six dimensions with anomaly free particle content. 

A particle spectrum, consisting of 9 tensor, 8 vector
and 20 hypermultiplets and $N=1$ supersymmetry
was obtained in six dimesnions 
by Sen\cite{sen} for $M$-Theory compactified on an orientifold of
$K_3\times S^1$. 
In this case, like in the ten dimensional one,
the twisted sector states can not be obtained by a direct
$M$-Theory calculation. However, Sen was able to determine the
spectrum in the twisted sector by comparing the nature of the
fixed points for $K_3\times S^1$ with those for $T^5$. It was
argued that the physics near the fixed points in the case of
$(K_3\times S^1)/ Z_2$ is identical to that for $T^5/Z_2$. As a
result, the contribution of the twisted sector states are
identical in the two cases.
This fact will also be utilized in our case below.
Since the extra $Z_2$'s that we apply 
act freely and keep the supersymmetry intact, 
the number of fixed points remains unchanged.
Their contribution
to the field content also does not change, since they still come
as tensor multiplets of the chiral $N=2$ algebra\cite{wit2,sen}
due to the supersymmetry preserving nature of the extra
projections\cite{chaudh}. 
This is further
confirmed by the fact that we are able to obtain anomaly free 
combination of fields in all our examples.

In this article, we present new examples of orientifold
compactifications of $M$-Theory by further orbifolding of the
models in \cite{sen} with respect to the finite abelian
automorphism groups of $K_3$\cite{walton}. 
In particular, we present several $N=1$
supersymmetric examples with different number of vector
multiplets. The orbifolds of $K_3$ for the case of type IIA
string compactification was discussed in \cite{scsen,chaudh}. 
These orbifolds provided examples of the dual pairs of the
heterotic string theories in dimensions six and less with
maximal supersymmetry, but with lower rank gauge groups. 
In this article we focus our attention on the $Z_2^k$ orbifolds
discussed in \cite{chaudh}. In our 
case, we combine one of these $Z_2$ actions with another
operation 
which changes the sign of the eleven dimensional 3-form fields
as well as the elven-dimensional coordinate $x^{10}$\cite{sen}. 
As a result we are
able to construct several models with $N=1$ supersymmetry, but with
different number of vector multiplets. Extra $Z_2$ symmetries,
in our case act freely. 
To avoid new fixed points, we combine these
$Z_2$'s with the translation symmetries along some of the
compactified directions. Consequently, like in \cite{chaudh}, our
result is strictly valid only in dimensions less than six.
The lower dimesnional spectra can however be seen to
arise from the corresponding six dimensional ones, with new
anomaly free combinations, in a straightforward way. 

We now begin by describing the $Z_2^k$ $(k=1, 2, 3, 4)$
orbifolds presented in \cite{chaudh}. The action of the
symmetries is represented by $k$ $Z_2$ generators, denoted by
$g_i$ $(i=1,..,k)$. Under these $Z_2$'s all the three self-dual two
forms and three of the nineteen anti-self-dual three forms are
invariant. On the remaining sixteen anti-self-dual two forms it
acts as:
\be
        g_1 : \left( (-1)^8, 1^8 \right),   \label{g1}
\ee
\be
        g_2 : \left( (-1)^4, 1^4, (-1)^4, 1^4 \right), \label{g2}
\ee
\be
        g_3 : \left( (-1)^2, 1^2, (-1)^2, 1^2, 
        (-1)^2, 1^2, (-1)^2, 1^2 \right), \label{g3}
\ee
\be
        g_4 : \left(-1, 1, -1, 1, -1, 1, -1, 1, -1, 1, 
                    -1, 1, -1, 1, -1, 1 \right), \label{g4}
\ee
with superscripts on 1 denoting the repeated entries. Individually,
as will be described below,
under any of these $Z_2$'s 34 of the 58 $K_3$ moduli fields
remain invariant. Similarly, 14 of the 22 two-forms on $K_3$ are
even under $Z_2$'s and 8 are odd. When more than one of these
discrete symmetries is used to mod out the original theory, the
resulting spectrum is determined by taking the intersection of
the individual ones. 

Above discrete symmetries are now used to obtain the untwisted
sector of the spectrum when $M$-Theory, with low energy spectrum
consisting of a graviton $G_{M N}$ and a third rank
antisymmetric tensor $A_{MNP}$ in eleven dimensions, 
is compactified on an orientifold of $K_3\times S^1$ . 
For this purpose, the first $Z_2$ in equation (\ref{g1}) 
is combined with an operation $A_{MNP} \rightarrow - A_{MNP}$
and $x^{10}\rightarrow -x^{10}$.
Here we have used the notation that $(M, N) = (0,...,10)$
and $(\mu, \nu) = (0,..,5)$. 
The rest of the $g_i$'s act in the same way as in
\cite{chaudh}. We would once again like to point out that 
projections in \cite{chaudh} with respect to $Z_2$'s in equations 
(\ref{g1})-(\ref{g4}) keeps the supersymmetries 
unbroken. As a result, further orbifolding, by $g_i$'s, of 
the resulting $N=1$ theory also keeps the supersymmetry intact. 

We now present the counting of the surviving degrees of
freedom when several $g_i$'s in
equations (\ref{g1})-(\ref{g4}) are applied on $G_{MN}$ and
$A_{MNP}$. First, the surviving degrees of freedom for $G_{MN}$ is
determined by examining the local structure of the moduli space.
Since the number of supersymmetries for the $Z_2$ projections 
by $g_i$'s 
remain unchanged, the local structure of the moduli space for the
compactifications on the above orbifolds of $K_3$ has the form
\cite{chaudh}:
\be
        {\cal M} = {{SO(20 - r, 4; R)}\over 
        {SO(20-r; R) \times SO(4;R)}},   \label{moduli}
\ee
where $r$ is the reduction in the rank of the gauge group 
from the maximal number 24, which is equal to the one 
for the toroidal
compactification of the heterotic string. Such reductions 
in the ranks
of the gauge group, for the orbifolds of $K_3$ described above,
can now be studied in the type IIA theory. We will use this
information to determine the number $r$ for 
the action of the various combinations of $g_i$'s.

All the 24 gauge fields in the $K_3$ compactification of type
IIA theory originate in the Ramond-Ramond (R-R) sector. 
Since for $K_3$, only nonzero Betti numbers are 
$b_0 = b_4 = 1$ and $b_2 =22$, one of the gauge fields in six
dimensions originates from its ten dimensional counterpart $A_M$.
Three-form field components $A_{\mu m n}$ give rise to 22 gauge
fields and the dualization of $A_{\mu \nu \rho}$ gives the
remaining one. Here $(m, n)$ denote the indices on $K_3$.
An observation of the form of $g_i$'s in
equations (\ref{g1})-(\ref{g4}) now gives the values:

\leftline{(i) $r = 8$ for the action of $g_1$,}
\leftline{(ii) $r = 12$ for the
action of $g_1$ and $g_2$ together,} 
\leftline{(iii) $r = 14$ for the action of 
$g_1$, $g_2$ and $g_3$, and finally} 
\leftline{(iv) $r= 15$, when all the
four $g_i$'s in equations (\ref{g1})-(\ref{g4}) are applied.}
The reduction in the number of gauge fields follows directly by
counting the number of 2-forms on $K_3$ that are projected out
by the action of $g_i$'s. We will use these informations to
determine the invariant components of the metric $G_{MN}$ under
compactification. 
The number of 2-form fields left invariant in the
cases (i)-(iv) above are respectively 
14, 10, 8 and 7. By
subtracting these numbers from the 
dimension of the coset (\ref{moduli}), and
taking into account the rank-reduction in the last paragraph, 
we get the number of invariant scalars from the $K_3$ part,
originating from the metric
$G_{MN}$ in ten dimesnions. These are respectively (i)
34, (ii) 22, (iii) 16 and (iv) 13. 

We now first construct an orientifold of $M$-Theory
for the action of 
symmetry $g_1$ and then obtain other models by additional
projections with respect to the remaining $g_i$'s. As mentioned
earlier, to avoid fixed points with respect to these additional
$Z_2$'s, one has to combine them with the translations on circles
\cite{scsen,chaudh}
by compactifying further to lower dimensions. However, since the
field content for the massless modes does not depend on these
shifts, these lower
dimesnional spectra follow directly from the toroidal
compactification of the six dimensional ones. Therefore the
effect of these extra $Z_2$'s can be seen directly in terms of a
six dimensional spectrum which we now present. 
In the next four paragraphs, we present
the field contents in the cases (i)-(iv) above,
starting from the fields $G_{MN}$ and $A_{MNP}$ in eleven
dimensions. We also show that they are all
consistent with the anomaly cancellation
requirements. 

(i) The projection with respect to $g_1$ leaves the components
$G_{\mu \nu}$ and $G_{(10) (10)}$ of $G_{MN}$ invariant,
and also gives 34 scalar fields from the local moduli
in $K_3$.  In addition, taking into account that $A_{MNP}$ is
odd under this $Z_2$ and $x^{10}$ changes sign, we find that
the component $A_{\mu \nu (10)}$ remains invariant and gives rise
to an antisymmetric tensor field in six dimensions. 
We also get 8 gauge fields
$A_{\mu m n}$ from the eight 2-forms of $K_3$ which are odd
under $g_1$. 14 more scalars arise from even 2-form
components $A_{m n (10)}$ on $K_3$. Together, these give us the
graviton, an antisymmetric tensor, 8 vectors and 49 scalars in
the untwisted sector\cite{sen}. 
Combining these with the states arising
from the twisted sectors, namely 8 tensor and 8 hypermultiplets,
we find the following final spectrum. (a) 9 tensor multiplets
(b) 1 graviton multiplet (c) 8 vector multiplets and (d)
20 hypermultiplets. These are precisely the combination needed
for the anomaly cancellatin in $N=1$ supersymmetric 
theories\cite{anomaly,sen}.

(ii) For this case the spectrum in the untwisted sector is
obtained by taking intersection with respect to the projections
in (i), as described in the last paragraph, with those with
respect to $g_2$. The surviving degrees of freedom from $G_{MN}$
now consists of $G_{\mu \nu}$, $G_{(10) (10)}$ and 22 local moduli
fields. Among the $A_{MNP}$ components, which survive this
projection are now (a) $A_{\mu \nu (10)}$, 
(b) 4 of the eight vectors 
$A_{\mu m n}$ in the last paragraph and 
(c) 10 of the fourteen scalars $A_{m n
(10)}$. Together these provide a graviton, an antisymmetric tensor,
4 vectors and 33 scalars in the untwisted sector. Combining
these, once again, with the fields from the twisted sectors,
namely the 8 tensor and 8 hypermultiplets we get the full field
content for this theory as: (a) 9 tensor multiplets, (b) a
gravity multiplet, (c) 4 vector multiplets and (d) 16
hypermultiplets. This is once again an anomaly free
spectrum\cite{anomaly}. As
stated earlier, $g_2$ acts freely only 
when it is combined with a half-shift along one of the circles
of compactification. For this purpose, one has to consider
this orientifold only in dimensions five. 
However, since the
field content for the massless fields does not depend on the
shift, the resulting five dimensional fields can
easily be seen to arise from the compactification of the
above massless spectrum in six dimensions on a circle. 
This is a new anomaly
free combination, other than the one discussed in \cite{sen}. 

(iii) The effect of the actions of $g_2$ and $g_3$ 
on the orientifold of $g_1$ in (i) can be
studied exactly in the same way as in (ii) above. The massless
fields from the untwisted sector now consist of a graviton,
1 antisymmetric tensor, 2 vector fields and 25 scalar fields.
Once again, combining these with the 8 tensor and 8
hypermultiplets of the twisted sector gives rise to 9 tensor
multiplets, a graviton multiplet, 2 vector multiplets and 14
hypermultiplets. This is an anomaly free field content. As in
the last paragraph, the two $g_i$'s have to be
combined with appropriate shifts along the compactified
directions. As a result, this time the model has a valid
interpretation only in four dimensions. However once again, 
the resulting four dimesnional fields can be seen to
originate from a six dimensional theory with the above anomaly
free field content through a simple toroidal compactification. 

(iv) Finally, when all the four $g_i$'s are applied together,
with $g_1$ acting as an orientifold as above, we have from the
untwisted sector, a graviton, 1 antisymmetric tensor, 1 vector
field and 21 scalars. Combining them with the twisted sector
states, i.e. 8 tensor and 8 hypermultiplets, we get another
anomaly free field content: 9 tensor multiplets, 1 gravity
multiplet, 1 vector multiplet and 13 hypermultiplets.

To conclude, we have presented new anomaly free
combinations in six dimensions that arise from 
the $K_3\times S^1$ orientifold of $M$-Theory by additional
projections with respect to the abelian automorphism groups of $K_3$.
One of the key ingredients in constructing these
$N=1$ supersymmetric theories has been the fact that the number of
supersymmetries do not reduce by the above actions of the
automorphisms of $K_3$. We
have also avoided the presence of any new twisted sector states
by combining the extra $Z_2$ projections with the shifts along
the directions on which the six dimesnional theory is
compactified. Another way to avoid additional fixed points may
be by combining these additional $Z_2$'s with the shifts along
the central charges of the supersymmetry algebra\cite{seva}.
We will then have a genuinely six dimensional theory with the
same field contents as mentioned in this paper. One may also directly
study the contributions from all the additional fixed
points, which have been avoided here, 
and see if the anomaly cancellation condition
is maintained.


\end{document}